\documentclass[aps,prc,onecolumn,superscriptaddress,showpacs]{revtex4-2}
\usepackage{graphicx}
\usepackage{bm}
\usepackage{multirow}
\usepackage{fontenc}
\usepackage{subcaption}
\usepackage{dcolumn}
\begin{document}

\title{Enhancing the performance of solenoidal spectrometers for inverse reactions}

\author{%%%% Author details
P. A. Butler }

%%%%%%%%% Insert author address here
\affiliation{Oliver Lodge Laboratory, University of Liverpool,
Liverpool L69 7ZE}

\begin{abstract}
Helical-orbit solenoidal spectrometers, in which the target and
detector are placed inside a uniform magnetic field, have been
utilised for more than a decade to study nuclear reactions in
inverse kinematics, induced by radioactive beams. Methods to improve
the final state energy resolution are presented and the inclusion of
an active gas target is proposed to improve the performance of the
spectrometer.
\end{abstract}

\maketitle
%% \linenumbers

%% main text
\section{Introduction}
\label{}  In order to understand the underlying quantum structure of
atomic nuclei, measurements that use reactions to probe nuclear
properties should be carried out with a precision that is sensitive
to this structure. Transfer reactions such as (d,p) and inelastic
scattering reactions such as (d,d') have been employed for many
decades to make direct measurements of the single-particle and
collective properties of nuclear states. For reactions induced by
intense beams of light ions bombarding stable targets, final-state
energy resolutions of $\approx 10$ keV can be achieved using
magnetic spectrographs, see, e.g.,~\cite{Von,Tho}. In order to study
nuclei beyond the line of stability, however, the detector system,
whose precision is determined by the detector design, the target and
the characteristics of the beam, has to be highly efficient in order
to exploit the much weaker flux of radioactive beams~\cite{Wim}. In
this case the states of interest are populated by an inverse
reaction, usually by bombarding a solid deuterated polyethylene
[(C$_2$D$_4$)$_n$] target. This would normally have the disadvantage
that the strong dependence of the energies of the emitted light ions
on their angle of emission will lead to kinematic broadening, and
the separation between the laboratory energies of the excited states
will be highly compressed. These difficulties have largely been
overcome by employing an ingenious technique, developed at the
Argonne National Laboratory, whereby the emitted light particles are
transported in a homogeneous magnetic field ${\bm B}$, parallel to
the beam axis, to a position-sensitive silicon-detector
array~\cite{Wuo,Lig}. Here both the particle energy and the distance
between the target and the intercept of the particle's trajectory
with the beam axis are recorded in order to determine the excitation
energy of the state of interest, which has a linear dependence on
these quantities (see section~\ref{energy_resolution}). This
spectrometer allows the angular distribution of the emitted
particles to be measured with high efficiency over a wide angular
range. In addition, the value of the ratio of mass-to-charge of the
particle can be determined from the time of flight, which is the
cyclotron period for the particle motion in a magnetic field. The
spectrometer employing this concept is called the Helical Orbit
Spectrometer (HELIOS) and has been used successfully for over a
decade at the ANL ATLAS facility. More recently the ISOLDE
Solenoidal Spectrometer, ISS, using the same principle of operation,
has also been in operation employing radioactive beams from the
HIE-ISOLDE facility, CERN, and the Solenoidal Spectrometer Apparatus
for Reaction Studies, SOLARIS~\cite{Kay}, has been constructed to
take beams from the FRIB facility in the Michigan State University
campus.

This paper presents some ideas for improving the response of
spectrometers using the HELIOS concept. In
section~\ref{energy_resolution} the various contributions to the
total energy resolution of the excited states are estimated and
suggestions for reducing some of these contributions are presented.
Section~\ref{active_target} presents a proposal to replace the solid
target by an extended gas volume, where the reaction vertex is
determined using a time projection chamber.

\section{Contributions to Energy Resolution}\label{energy_resolution}
We consider the 2-body reaction $M_1 + M_2 \rightarrow M_3 + M_4$
where $M_1$ is the mass of the (heavy) projectile and $M_2$ is the
mass of the (light) target.  For the spectrometer using the HELIOS
concept, the light ejectile, having a charge state $q$ and mass
$M_4$ is emitted at the origin where the reaction occurs, the
vertex. The particle, having energy $E_4$ in the laboratory frame of
reference, travels in a magnetic field with flux density $\bm{B}$
which is parallel to the beam axis until its trajectory intercepts
with the beam axis at distance $z$ from the origin. In this case the
energy of the excited state of the residual nucleus of mass $M_3$
can be determined from the reaction $Q$-value, which is a linear
function of $E_4$ and $z$:
\begin{equation}
\label{Qequation}
Q = a E_4 + b - c z
\end{equation}

where $a,b,c$ are constants given by

 \begin{displaymath}
a = (M_3 + M_4)/M_3
\end{displaymath}

 \begin{displaymath}
b =  a M_4  v_{cm}^2 /2 - M_2 E_0 / (M_1 + M_2)
\end{displaymath}

and

 \begin{displaymath}
c = a M_4 v_{cm}/ T_{cyc}
\end{displaymath}

In the aforementioned expressions $E_0$ is the beam energy, $v_{cm}$
is the velocity of the center-of-mass system,

 \begin{displaymath}
v_{cm} =  \sqrt{2 M_1 E_0}/(M_1 + M_2)
\end{displaymath}

and $T_{cyc}$ is the cyclotron period of the light particle in the magnetic field,

 \begin{displaymath}
T_{cyc} = 2 \pi M_4 / (q e B)
\end{displaymath}

In the case of inelastic scattering reactions for which $M_2 = M_4$,
$b = 0$.

\begin{figure}[htb]
\hspace{-0.5cm}
\begin{subfigure}{0.35\textwidth}
\includegraphics[width=49mm]{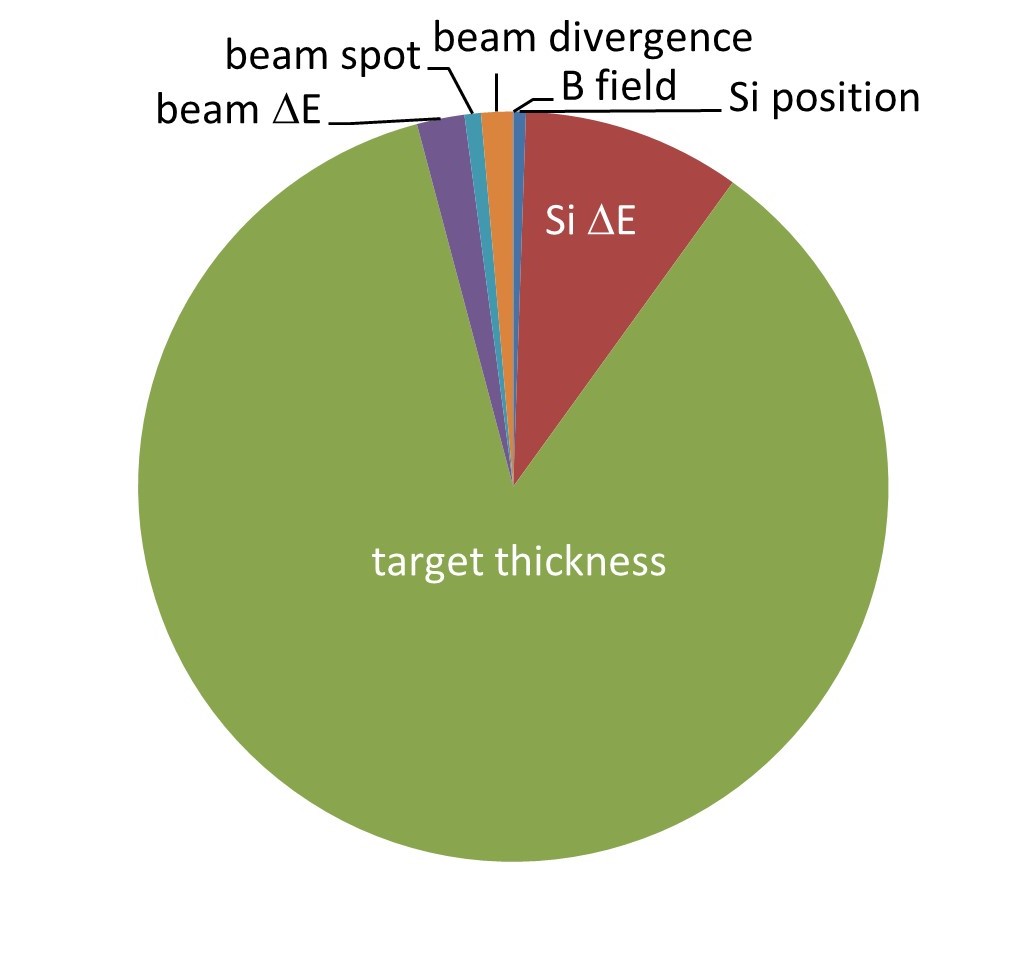}
\end{subfigure}%
%\hspace{-1cm}
\begin{subfigure}{0.35\textwidth}
\includegraphics[width=49mm]{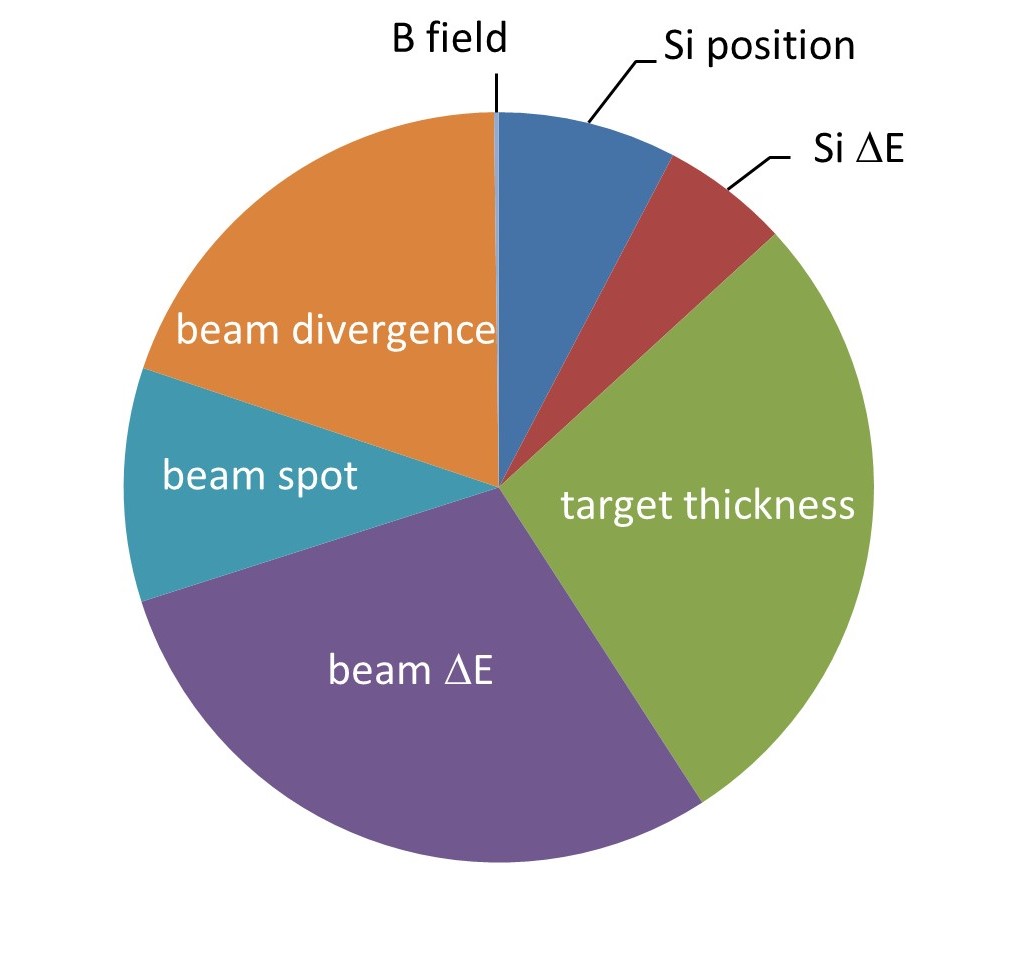}
\end{subfigure}%
%\hspace{1cm}
\begin{subfigure}{0.35\textwidth}
\includegraphics[width=49mm]{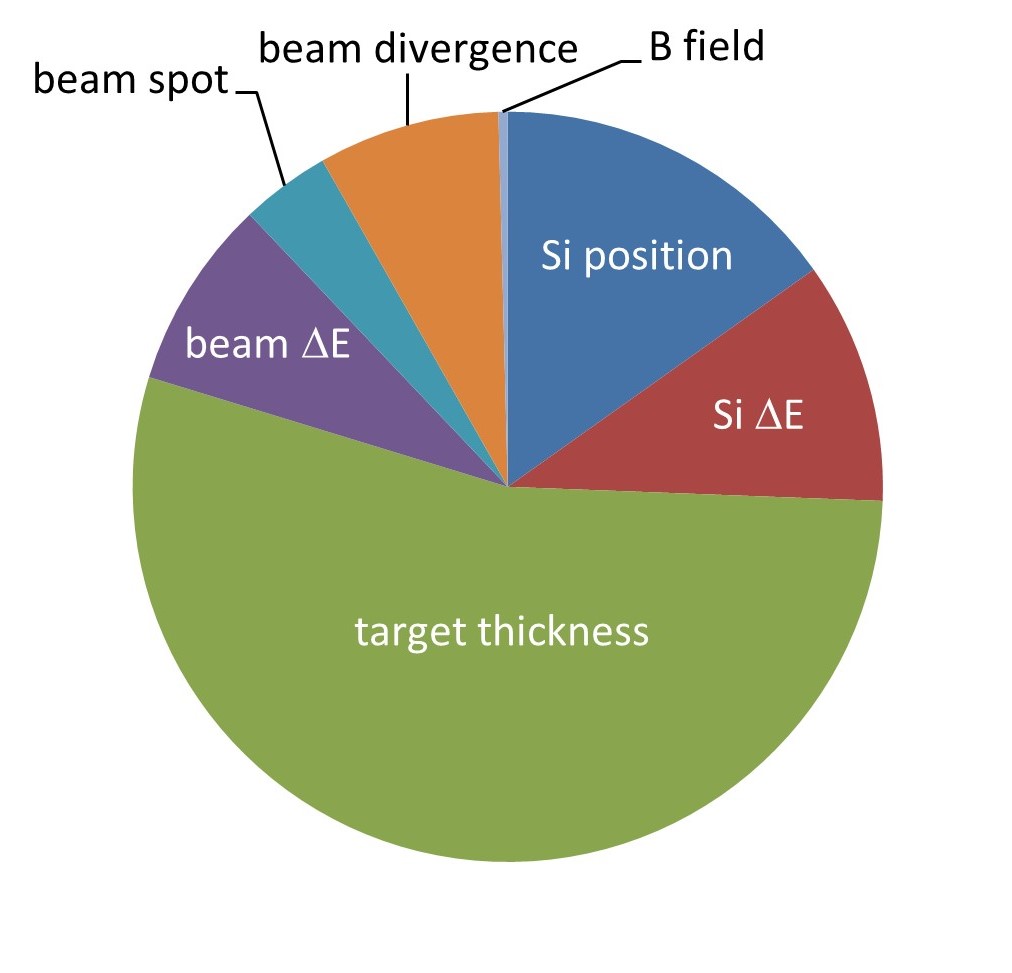}
\end{subfigure}
\caption{The contributions to the total $Q$-value resolution of the
ISOLDE Solenoidal Spectrometer for differing conditions. The left
figure assumes that the uncertainties are given by set 1 in
table~\ref{uncertainties} that are the typical values for HELIOS
spectrometers. For the central figure the intrinsic energy
resolution is reduced from 50keV to 10keV FWHM and the target
thickness is reduced from 165 to 10$\mu$g/cm$^2$. For the right
figure the beam characteristics are optimised by reducing the
transverse emittance and energy spread. The area of each sector is
proportional to the square of the $Q$-value resolution arising from
the contribution alone.}
 \label{Qvalres}
\end{figure}

To determine the various contributions to the spread in the measured
values of $Q$, the reaction and trajectory from the target to the
detector can be simulated under differing conditions. In this Monte
Carlo simulation~\cite{ButGH}, the centre-of-mass scattering angle
was stepped over in $0.2^{\circ}$ increments and 100 passes made for
each angle. For each pass the beam trajectory, ejectile trajectory
and detector response were calculated assuming random (usually
Gaussian) distributions of position, angle, energy etc.; the
ejectile trajectory was calculated by integrating the equations of
motion in the magnetic and electric fields using a step time of 1ps.
The incident beam will have a divergence and finite size at the
target, as well as a spread in energy. Both the heavy beam and the
emitted light particle will suffer energy loss, energy straggling
and multiple scattering in the target. In the passage to the
detector the magnetic field will not be perfectly homogeneous. When
the light particle reaches the silicon detector there will be an
uncertainty in the measured value of $z$, the distance between the
intercept of its trajectory on the beam axis and the target. These
simulations modelled the response of the Si array for the ISS, which
has the shape of a hexagonal prism in which each of the six
rectangular faces are parallel to the beam direction and consist of
four 22mm x 125mm double-sided silicon detectors. The perpendicular
distance from the face of each detector to the beam axis is $\approx
29$mm so that the detectors subtend $\approx 70\%$ of $2 \pi$ in the
plane transverse to this axis. The Si detectors are segmented along
the $z$-direction to give $z_{Si}$ and along the transverse
direction, enabling the radius of the trajectory at the intercept
$r_{Si}$ to be determined. As the Si array is at a finite radius
from the beam axis these values are used to extrapolate the
particle's trajectory to the beam axis. The quantities $z, r, z_{\rm
Si}, r_{\rm Si}$ are defined in the appendix that outlines a simple
algorithm that can be used to determine $z$ from $z_{\rm Si}$ and
$r_{\rm Si}$. The finite pitch of the segmentation will give rise to
an uncertainty in the value of $z$ and hence the $Q$-value. Finally,
the intrinsic energy resolution of the Si array will also contribute
to the uncertainty in the $Q$-value. For a typical case, the
Q-value, calculated from the values of $z$ and $E_4$ determined for
each event, was sampled several thousand times, allowing its
variance to be estimated.

The parameters used in the simulations are given in
table~\ref{uncertainties}. The energy loss and energy straggling of
the beam and light ions in the target were calculated using the SRIM
codes~\cite{Zie}. The angular dependence of the multiple scattering
was assumed to have a Gaussian distribution whose width was
calculated using the Fano prescription described by
Kantele~\cite{Kan}. For traversing through gas (see
section~\ref{active_target}) the value of the width was reduced in
order to empirically reproduce the measurements of Kuhn {\it et
al.}~\cite{Kuh}.
\begin{table}[h]
\centering \caption{\label{uncertainties} The various contributions
to the uncertainty in the value of $Q$. All uncertainties are FWHM
except for the detector pitch and target thickness.}
\begin{center}
\begin{tabular}{|l|c|c|c|}
\hline
source of uncertainty & \hspace{0.5cm} set 1 \hspace{0.5cm} & \hspace{0.5cm} set 2 \hspace{0.5cm} & \hspace{0.5cm} set 3 \hspace{0.5cm} \\
\hline
Si pitch $z$ direction(mm)  & 0.95 & 0.95 & 0.95\\
Si pitch transverse direction (mm)  & 2 & 2 & 2 \\
detector $E$ resolution (keV) & 50 & 10 & 10\\
target thickness (mg/cm$^2$) & 0.165 & 0.010 & 0.010\\
beam $E$ spread (\%) & 0.4 & 0.4 & 0.15 \\
beam spot (mm) & 2.3 & 2.3 & 1.0\\
beam divergence (mrad) & 1.8 & 1.8 & 0.8 \\
$B$ variation (\%) & 0.5 & 0.5 & 0.5\\
\hline overall $Q$-value resolution (keV) & 160 & 45 & 30 \\ \hline
\end{tabular}
\end{center}
\end{table}
Figure~\ref{Qvalres} shows how the contributions to the expected
$Q$-value resolution for the inverse reaction
$^{206}$Hg(d,p)$^{207}$Hg at a bombarding energy of 7.38MeV/u. The
nominal $Q$-value for the reaction was set to zero. The left figure
assumes that the uncertainties are given by set 1 in
table~\ref{uncertainties} that are the typical values for HELIOS
spectrometers. The full width at half maximum (FWHM) Q-value
resolution actually measured using the HELIOS Si array,
$165\mu$g/cm$^2$ (C$_2$D$_4$)$_n$ target and a $^{206}$Hg beam from
the HIE-ISOLDE post-accelerator was 140keV~\cite{Tan} (the
characteristics of the Si array used in this experiment are given in
ref~\cite{Lig} and differ slightly from those given in set 1). The
central figure assumes the values given in set 2 where the intrinsic
energy resolution is reduced from 50keV to 10keV FWHM and the target
thickness is reduced from 165 to 10$\mu$g/cm$^2$. This improves the
overall $Q$-value resolution from $\approx 160$ to $\approx 45$keV
FWHM. The loss in luminosity can be regained to some extent by
employing multiple targets, spaced apart sufficiently to identify
the reaction vertex by measuring the heavy recoil's time-of-flight
to a detector placed downstream. An intrinsic Si energy resolution
of 10keV FWHM or better should be achievable for proton
detection~\cite{Ste} although for the complex Si arrays used in the
HELIOS set-up the best resolution achieved at present is 25keV
FWHM~\cite{Lig}. As can be seen in figure~\ref{Qvalres} significant
contributions to the overall $Q$-value resolution come from the beam
characteristics, in this case being typical values for
HIE-ISOLDE~\cite{Kad}. The transverse emittance of the beam, that
determines the beam divergence and size of the beam spot, can be
reduced by collimation while the energy spread can be reduced by
operating a Linac cavity of the post-accelerator as a
buncher~\cite{Loz}. Such improvements in the beam characteristics
can also be achieved by injection into, and extraction from, a
storage ring that cools the circulating beam~\cite{But}. The values
given in set 3 can be obtained for optimised beams from HIE-ISOLDE.
The contributions to the $Q$-value resolution, $\approx 30$keV FWHM
overall, are shown in figure~\ref{Qvalres}. However, in order to
reduce the transverse emittance to a value that gives the tabulated
beam characteristics, the beam intensity will be reduced to $\approx
25$\% of its initial value. The reduction in beam intensity and
target thickness would prohibit the use of radioactive beams in many
cases.

\begin{figure}[htb]
\vspace{9pt}
\includegraphics[width=180mm]{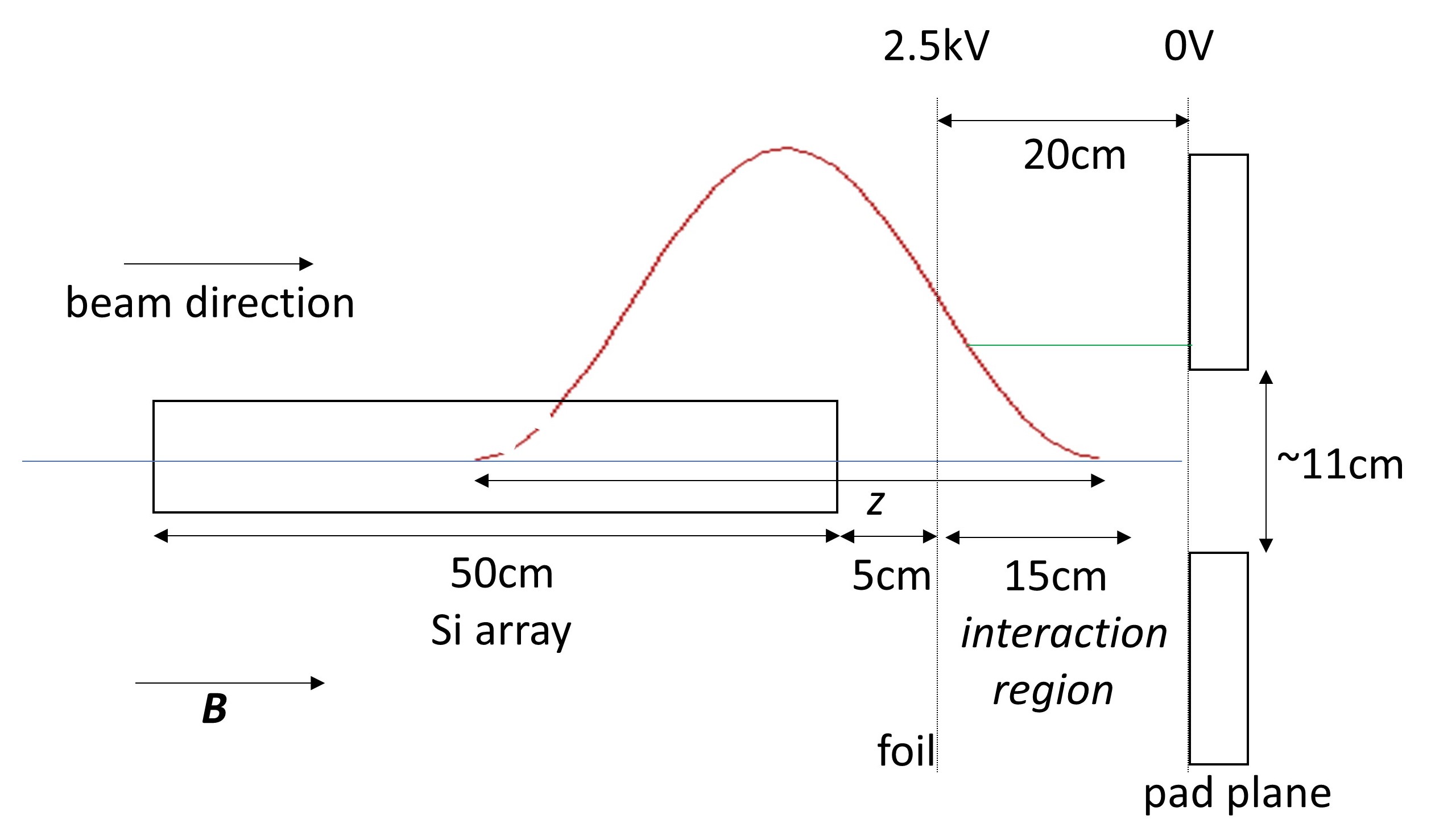}
\caption{Layout of the hybrid detector, HELIOS-TPC. The ${\bm E}$
field is parallel to the ${\bm B}$ field and the beam direction. The
interaction region is 15cm in the $z$ direction.} \label{layout}
\end{figure}

\section{Adoption of gas active target, HELIOS-TPC}\label{active_target}

In this mode of operation the target is now a gas, e.g., deuterium,
in a homogeneous electric field ${\bm E}$, parallel to ${\bm B}$.
This combines the HELIOS concept with that of the Time-Charge
Projection chamber (TPC), the latter employed successfully in
low-energy nuclear physics applications for two decades (see
e.g.~\cite{Miz,Dem}) and more recently operating inside a solenoidal
magnetic field~\cite{Bra, Pol}. For recent reviews of active
targets, see ~\cite{Bec,Baz}.  In an active target pure gases such
as H$_2$, D$_2$, N$_2$ can be employed, for which there will be
almost no background from carbon-induced events. For $^{3,4}$He a
small amount (5\%) of N$_2$ or CO$_2$ is added. For rare,
radioactive isotopes such as tritium, self-contained and sealed
cells can be used to reduce the activity from the gas volume, as
suggested in ~\cite{Ayy}. The proposed layout of the hybrid
spectrometer is shown in figure~\ref{layout}. Here, the high voltage
plane lies close to the Si array, and the ionisation charges in the
gas produced by the passage of the light ejectile through the gas
are collected in the sensor plane that is placed downstream of the
interaction region. For each event the values of $z_{\rm p}$ and
$r_{\rm p}$ (see appendix) are determined from the drift time and
position of the collected charges on the sensor plane. The $z$
position of the reaction vertex can then be deduced by extrapolation
using the algorithm described in the appendix. The sensor plane is
typically equipped with a Micromegas device~\cite{Gio}, consisting
of a printed circuit board covered with electrodes, or {\it pads}.
The charge drift velocity in the arrangement shown in
figure~\ref{layout} will give a maximum collection time of $\approx
10\mu$s for a pressure of 50Torr D$_2$.
\begin{figure}[htb]
\begin{subfigure}{0.5\textwidth}
\includegraphics[width=100mm]{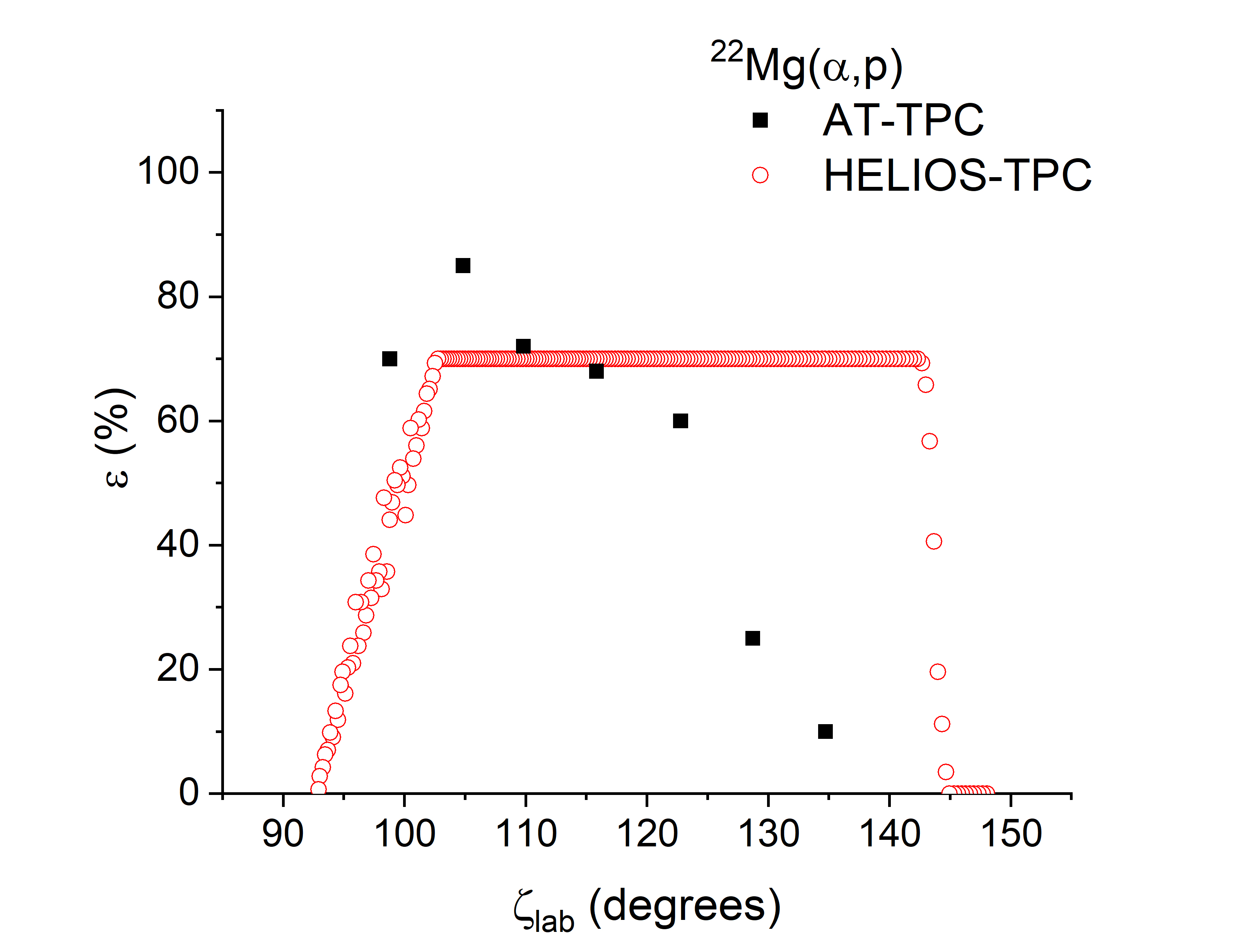}
\end{subfigure}%
\begin{subfigure}{0.5\textwidth}
\includegraphics[width=100mm]{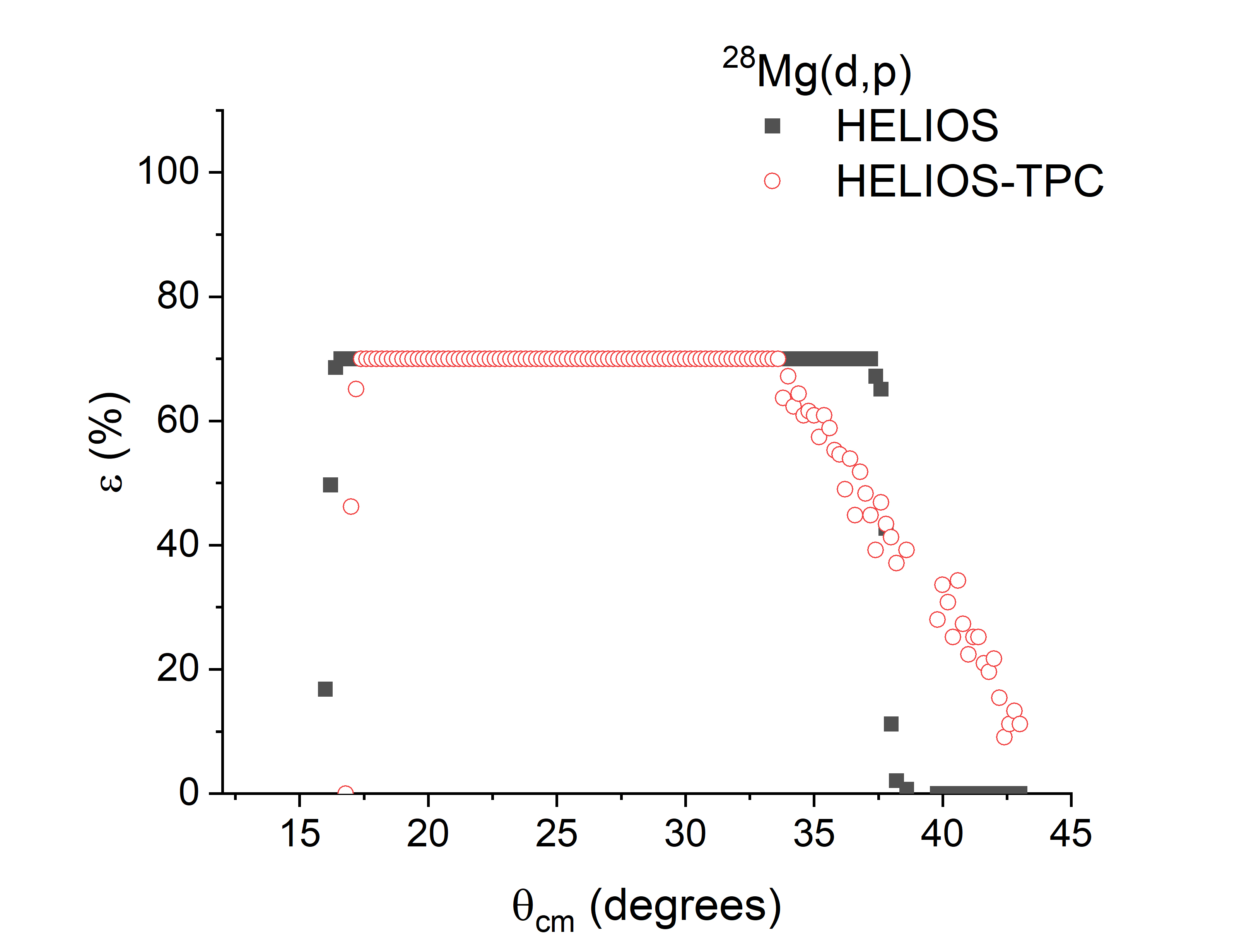}
\end{subfigure}
\begin{subfigure}{0.5\textwidth}
\includegraphics[width=100mm]{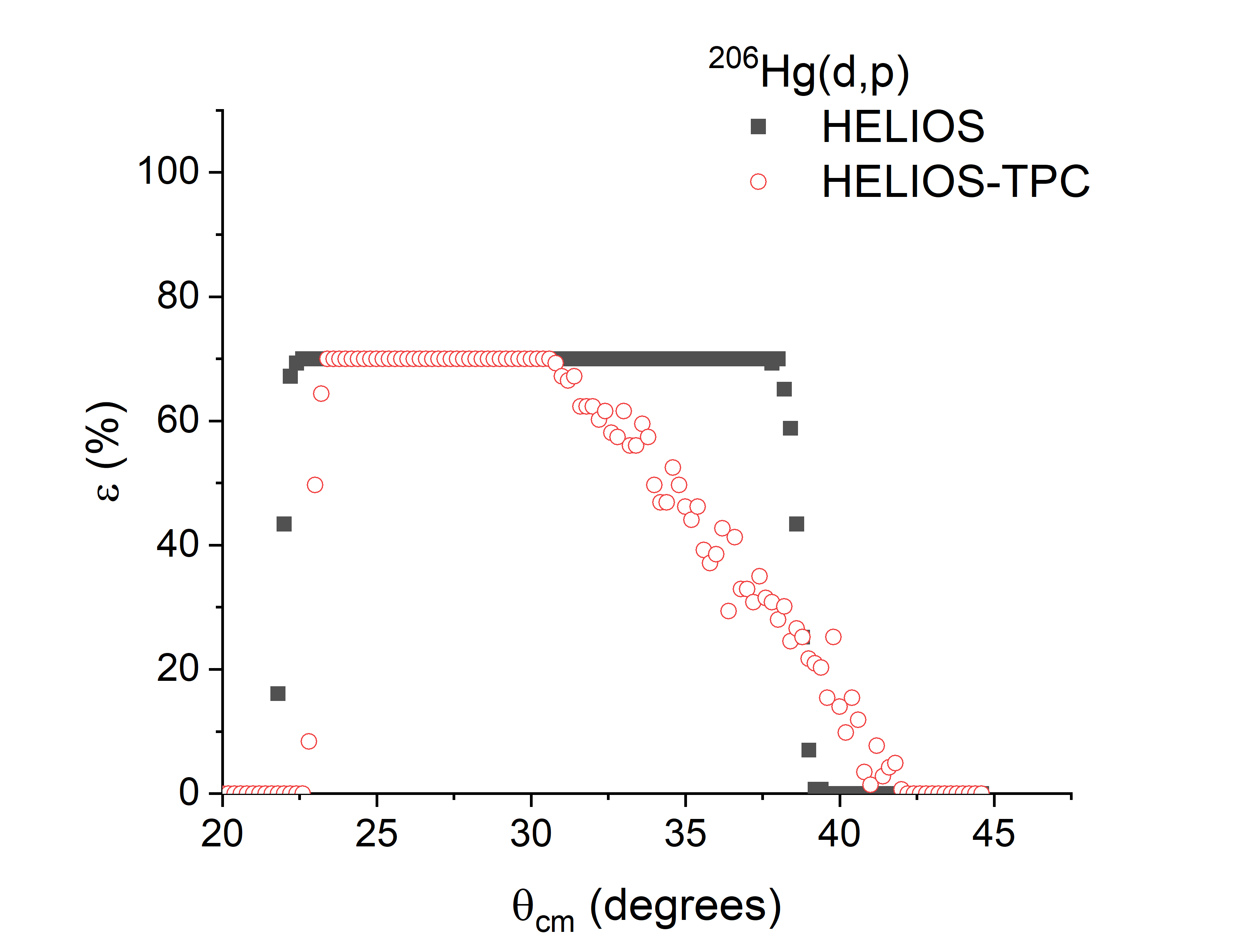}
\end{subfigure}%
\begin{subfigure}{0.5\textwidth}
\includegraphics[width=100mm]{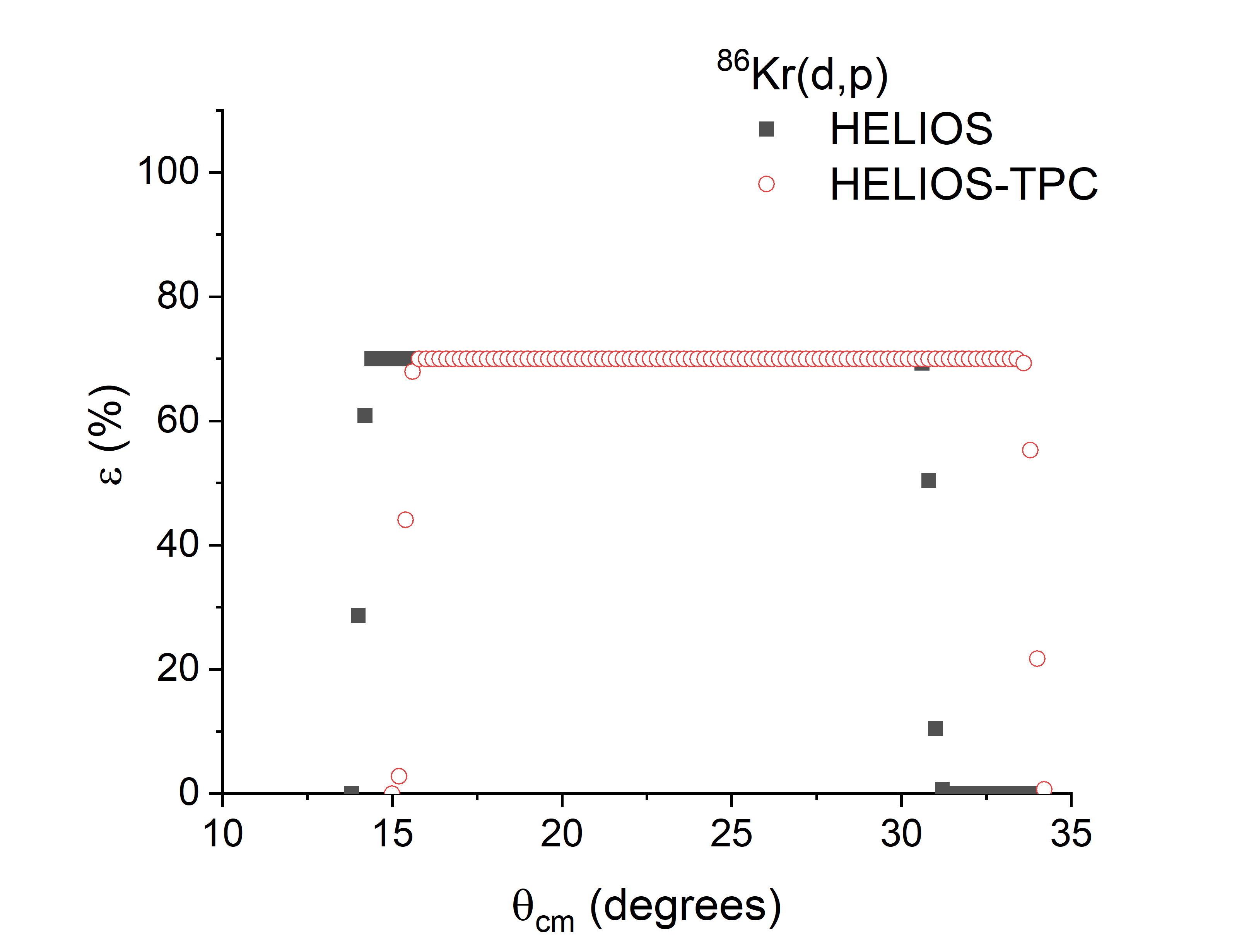}
\end{subfigure}
\caption{Efficiency, $\epsilon$, for detection of protons in the
silicon array from four reactions listed in table~\ref{performance}
as a function of the laboratory scattering angle of the ejectile
$\zeta_{\rm lab}$ or centre-of-mass scattering angle $\theta_{\rm
cm}$. For the $^{22}$Mg($\alpha$,p)$^{25}$Al reaction the black
solid squares are measured values for the AT-TPC~\cite{Ran}. For the
(d,p) reactions the black solid squares are values from the
simulations described in the text for the HELIOS-like spectrometer
ISS. The red open circles are for the hybrid HELIOS-TPC spectrometer
assuming the layout of figure~\ref{layout} with 50Torr He or D$_2$
gas. The simulations use a nominal value for the reaction $Q$ of
zero.} \label{efficiency}
\end{figure}
The reaction zone and the immediate surrounding volume are shielded
from the time-projection pads. This ensures that the beam particles
and small angle elastic scattering events are not detected,
important to maximise the luminosity of the experiment (see later).
Similar considerations led to the design of the TACTIC~\cite{Fox}
and ANASEN~\cite{Kos} active-target detectors. In order to
investigate the performance of this hybrid spectrometer simulations
were carried out assuming the same silicon array as described in
section~\ref{energy_resolution} and 3mm time-projection pads at a
radius of 5.5cm. For reactions such as (d,p) or ($\alpha$,p) where
protons emitted backwards are detected, events arising from elastic
scattering can be easily rejected. For forward-going reactions such
as (d,d') the pads are placed upstream of the target region. In this
case, the reaction vertex, pad and Si detector define the cyclotron
trajectory in the x-y plane, which is then matched to that
calculated from $E_4$ and $z$. The perturbation of this trajectory
due to energy loss in deuterium or helium gas is very small, and can
be corrected.

In table~\ref{performance} the maximum luminosity $L_{max}$, defined
numerically as the product of the maximum beam intensity $I$
(ions/s), the target thickness $t$ (mg/cm$^2$) and the absolute
efficiency of the detector system $\epsilon$ is estimated for the
hybrid spectrometer and compared with experimental measurements
using radioactive beams. The measurements used the active-target
spectrometer AT-TPC with 600Torr He (with 5\% CO$_2$) to study the
reaction $^{22}$Mg($\alpha$,p)$^{25}$Al~\cite{Ran} or ISS with
(C$_2$D$_4$)$_n$ targets to study $^{28}$Mg(d,p)$^{29}$Mg~\cite{Mac}
and $^{206}$Hg(d,p)$^{207}$Hg~\cite{Tan}. For the hybrid system the
beam interaction region is assumed to be 15cm of 50Torr He or 50Torr
D$_2$. The latter is equivalent to 0.67mg/cm$^2$ (C$_2$D$_4$)$_n$.
An important consideration in determining $L_{max}$ is the rate of
elastically scattered target recoils that can potentially give a
high instantaneous rate in the pad detectors, limiting the maximum
beam intensity.  In the case of HIE-ISOLDE the magnesium beam
intensities given in table~\ref{performance} for the simulations are
limited by the capability of the ISOLDE primary target ion-source.
For the $^{206}$Hg beam intensity given in the table the average
rate of scattered target recoils is $10^3$/s integrated over all
angles for which the trajectory of the ion in the magnetic field,
moderated by energy loss in the gas, reaches further than 5.5cm from
the beam axis. The macroscopic beam structure of the HIE-ISOLDE
linear accelerator has $\approx 1$ms pulses with a repetition rate
determined by the charge breeding time in the EBIS ion source. For
mass $\approx 200$ ions this is $\approx 200$ms~\cite{Kad}, giving
an instantaneous rate of $\approx 200$ per beam pulse, i.e 2 events
in the $10 \mu$s collection time. While this is approaching the
limit for the sampling electronics for a full tracking active-target
spectrometer such as the AT-TPC~\cite{Bra,Bec}, the simplified pad
array envisaged here should be able to accept a higher event rate.

The angular-dependent efficiency of the hybrid system that takes
into account the electric field cage surrounding the active target
region is shown in figure~\ref{efficiency} for four reactions listed
in table~\ref{performance} and compared with the simulated
efficiency of HELIOS employing a (C$_2$D$_4$)$_n$ target or the
measured AT-TPC efficiency~\cite{Ran}. The cut-off at low values of
$\theta_{\rm cm}$ (high values of $\zeta_{\rm lab}$) is determined
by the chosen energy threshold for detection of protons in the Si
array. The highest value of $\theta_{\rm cm}$ detected is limited
either by the closest distance of the interaction region to the Si
array or by the finite radius of the magnet. The overall efficiency
is 50 - 60\%, comparable to that of the ISS and that of the AT-TPC.
Table~\ref{performance} demonstrates that the maximum luminosity for
the hybrid spectrometer compares favourably with both active target
and HELIOS spectrometers. Also included in table~\ref{performance}
are the luminosities estimated for the $^{86}$Kr(d,p)$^{87}$Kr and
$^{136}$Xe(d,p)$^{137}$Xe stable-beam reactions. These reactions
were studied using HELIOS at the ATLAS facility~\cite{Sha,Kay2}
where the intensity of the beam, produced by an ECR source with
100\% duty factor, was limited to prevent damage of the targets. For
the hybrid spectrometer, it was assumed that the beam intensity is
limited by the instantaneous count rate arising from the poor duty
factor of the HIE-ISOLDE linac. In this case the luminosity for the
hybrid device is within a factor of two of that achieved using
HELIOS.

\begin{table*}[h]
%\footnotesize
%\centering
\caption{\label{performance} Maximum luminosities and widths of
$Q$-value distribution estimated for various reactions from the
simulations of the performance of the HELIOS-TPC hybrid detector
with 50Torr gas.  These are compared with the actual luminosities
achieved for the experiments (see the text) and experimental
measurements of the $Q$-value resolution. For the simulations the
nominal value of the reaction $Q$ was zero. For the experiments
using (C$_2$D$_4$)$_n$ targets, the total target thickness is given.
}
\begin{tabular}{|c|c|c|c|c|c|c|c|c|c|}
\hline
\multicolumn{1}{|c|} { } & \multicolumn{4}{|c|} { } & \multicolumn{5}{|c|} { } \\
\multicolumn{1}{|c|} { } & \multicolumn{4}{|c|} {  Hybrid HELIOS-TPC
(simulation)} & \multicolumn{5} {|c|} { AT-TPC or ISS (see text)}
\\
\multicolumn{1}{|c|} { } & \multicolumn{4}{|c|} { } & \multicolumn{5}{|c|} { } \\
\hline
 & beam & beam & luminosity & Q-value & beam & target &
luminosity & Q-value & \\
 reaction &  energy   & intensity & $L_{max}$  & FWHM & intensity &
thickness & $L_{max}$ &  FWHM  & reference\\
& MeV/u & ions/s & $I \cdot t \cdot \epsilon$ & keV & ions/s & mg/cm$^{-2}$ & $I \cdot t \cdot \epsilon$ &  keV & \\
 \hline & & & & & & & & & \\
 $^{22}$Mg($\alpha$,p)  &  5 &  $6 \cdot
10^5$ & $6 \cdot 10^4$ & 93 &
900 & 13 & $6.5 \cdot 10^3$ & $\approx 300$ & \cite{Ran} \\
 & & & & & & & & & \\
\hline  & & & & & & & & & \\
$^{28}$Mg(d,p) &  9.47  & $10^6$ & $9.7 \cdot 10^4$ & 75 &
$10^6$ & $0.12$ & $1.8 \cdot 10^4$ & $ 130$ & \cite{Mac} \\
 & & & & & & & & & \\
\hline  & & & & & & & & & \\
$^{206}$Hg(d,p) & 7.38 & $1.5 \cdot 10^5$ & $1.3 \cdot 10^4$ &
70 & $5 \cdot 10^5$ & 0.17 & $1.2 \cdot 10^4$ & $ 140$ & \cite{Tan} \\
 & & & & & & & & & \\
\hline & & & & & & & & & \\
$^{86}$Kr(d,p) & 10 & $2.5 \cdot 10^6$ & $2.7 \cdot 10^5$ &
75 & $5 \cdot 10^7$ & 0.06 & $4.3 \cdot 10^5$ & $ 80$ & \cite{Sha} \\
 & & & & & & & & & \\
\hline & & & & & & & & & \\
$^{136}$Xe(d,p) & 10 & $5 \cdot 10^5$ & $5 \cdot 10^4$ &
75 & $5 \cdot 10^6$ & 0.15 & $1.3 \cdot 10^5$ & $ 100$ & \cite{Kay2} \\
 & & & & & & & & & \\
\hline
\end{tabular}
%\end{adjustwidth}
\end{table*}

Another advantage of employing an active gas target is that the
energy loss of the beam in the target and the energy loss of the
ejectile in the gas volume and any foils (typically 100keV or less)
can be determined reasonably accurately for each event.  This means
that the only contribution from the gas to the overall energy
resolution arises from multiple scattering and straggling. For a
pressure of 50Torr gas the total FWHM $Q$-value resolution is
estimated for the five reactions listed in table~\ref{performance}.
In these simulations it is assumed that the FWHM uncertainty in the
$z$-position of the ejectile trajectory in the gas from the charge
collection time is $1$mm. The goal of achieving this position
resolution will determine the design of the pad structure. In
addition, beam energy loss, straggling and multiple scattering in
the gas entrance window (assumed to be $100\mu$g/cm$^2$
polypropylene) and the path to the interaction region were taken
into account. The characteristics of the Si detector array, the
beam, and the magnetic field were assumed to be the same as those
listed in set 1 in table~\ref{uncertainties}. The energy loss of the
ejectiles in the foil that provides the equipotential plane near the
Si array, assumed to be metal-coated $50\mu$g/cm$^2$ polypropylene,
was ignored: for 5MeV protons the energy loss in such a foil is
$\approx 5$keV that can be estimated for each event. The values of
the $Q$-value resolution determined from the simulations are given
table~\ref{performance}. In the case of the radioactive beam
reactions these values are significantly smaller than those measured
for the active target spectrometer~\cite{Ran} or the ISS
spectrometer~\cite{Mac,Tan}. For the stable beam reactions
$^{86}$Kr(d,p) and $^{136}$Xe(d,p) the $Q$-value resolution was
estimated to be 75keV for the hybrid spectrometer. This can be
compared to the 80keV value measured for the former reaction using
HELIOS with a $60\mu$g/cm$^2$ target~\cite{Sha} and 100keV for the
latter reaction using a $150\mu$g/cm$^2$ target~\cite{Kay2}.
However, an improved energy resolution can be achieved if the gas
pressure is reduced: for a D$_2$ pressure of 10Torr and intrinsic Si
energy resolution of 20keV the $Q$-value resolution is estimated to
be 45keV for the $^{206}$Hg(d,p)$^{207}$Hg reaction and 55keV for
the $^{86}$Kr(d,p)$^{87}$Kr and $^{136}$Xe(d,p)$^{137}$Xe reactions.
In this case the maximum luminosity is reduced by between third and
a half, and remains comparable to that measured using the HELIOS
spectrometers.

\section{Summary}
The various contributions to the final-state, or $Q$-value energy
resolution of a helical-orbit solenoidal spectrometer have been
considered. The largest contribution usually arises from effects of
energy loss and multiple scattering in the target, with significant
contributions from intrinsic energy resolution of the Si detector
array and the characteristics of the beam. The $Q$-value energy
resolution can be improved by reducing the target thickness and
manipulating the beam, but the resulting large reduction in
luminosity may not be desirable for studying reactions induced by
low-intensity radioactive beams. It is proposed to replace the
composite solid target by an extended gas volume, e.g. pure
deuterium, so that the hybrid spectrometer combines the properties
of the HELIOS spectrometer and that of a time-projection chamber.
This will remove contributions from carbon-induced reactions
involving hydrogen or deuterium targets and allow helium-induced
reactions to be studied. It is shown that HELIOS-TPC will enable
experiments to achieve good energy resolution without compromising
the luminosity of the experiment.

\section*{Appendix: Algorithm to extrapolate intersection of ejectile trajectory on beam axis}
\begin{figure}[htb]
%\vspace{9pt}
\centering
\includegraphics[width=125mm]{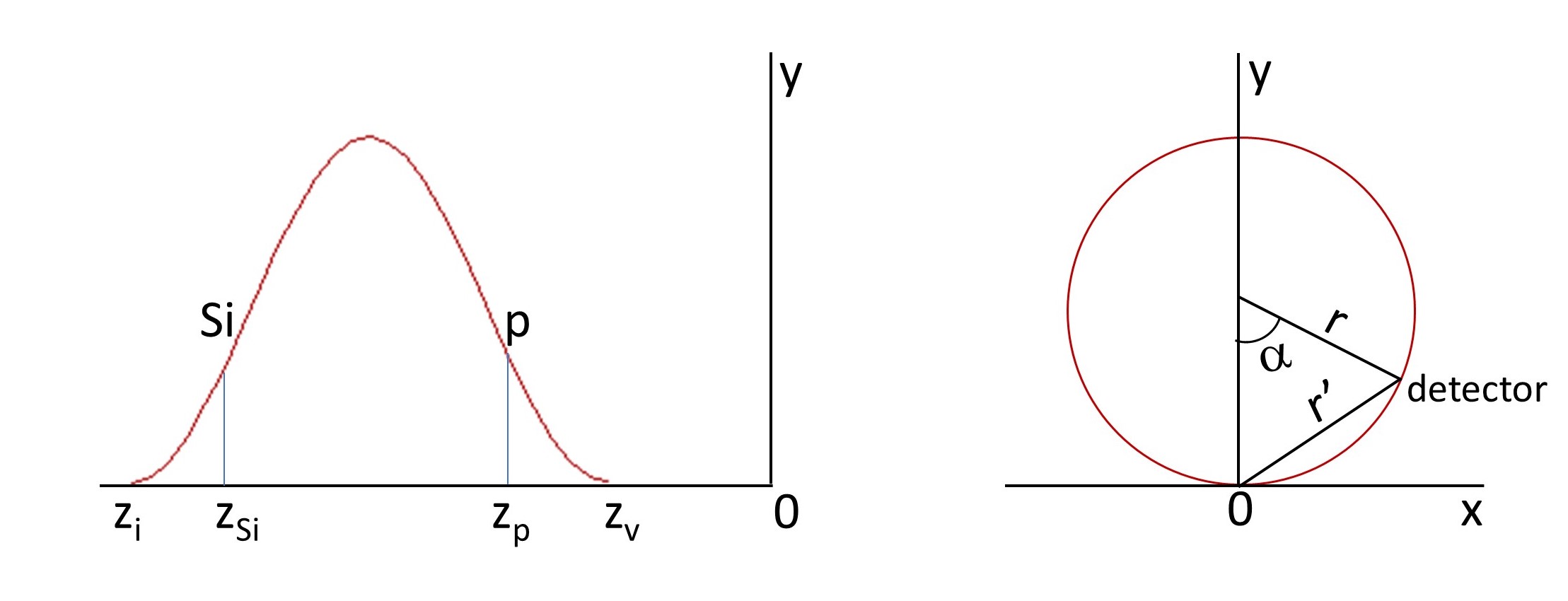}
\caption{The projection of the trajectory of the light particle in
the magnetic field in the $z-y$ and $x-y$ planes, showing its
relation to the position of the intersection on the pad detector
($z_p$) and silicon detector ($z_{Si}$).} \label{algorithm}
\end{figure}
The following assumes an active target configuration as shown in
figure~\ref{layout}, where the reaction vertex is determined using
the time projection method.  If a solid (C$_2$D$_4$)$_n$ target is
employed then only the second step is necessary, with $z_v = 0$.

\vspace{1cm}

{\bf First step.}  The value of $z_v$, the position of the reaction
vertex, is extrapolated from $z_p$, the position of the ejectile
trajectory measured at radius $r_p$ using the pad detector. The
value of $r_p$ is related to the rotation of the cyclotron motion in
the magnetic field, see figure~\ref{algorithm}, by

\begin{displaymath}
r' = r_p = \rho_{max} \sin( \alpha_p /2)
\end{displaymath}

where

\begin{equation}
\rho_{max} = 2r = 2 \frac{\sqrt{2 E_4 M_4}}{q e B } \sin\zeta
\label{equ:rhomax}
\end{equation}

and

\begin{equation}
\cos\zeta = (z_i - z_v)/  \frac{2 \pi \sqrt{2 E_4 M_4}}{q e B }
\label{equ:cos_zeta}
\end{equation}

In these equations $\zeta$ is the laboratory angle of emission of
the ejectile, with respect to the beam axis and the direction of
${\bm B}$. The value of $z = z_i - z_v$ ($z_i$ is the position of
intersection on the beam axis after one cyclotron period) can be
estimated using equation~\ref{Qequation}, since $E_4$ is measured.
As $z_i - z_v \gg z_p - z_v$ a nominal value of $Q$ can be used to
make an initial estimate of $z$ and hence $\alpha_p$.

Now,

\begin{displaymath}
(z_p-z_v)/(z_i -z_v) = \alpha_p/2\pi
\end{displaymath}

allowing $z_v$ to be determined.

\vspace{1cm}

{\bf Second step.} The value of $z_i$ is estimated more accurately
from $z_{Si}$, which is measured using the Si detector array. The
procedure is the same as for step 1 except that the radial position
at the Si detector (from the transverse displacement and detector
geometry) is given by

\begin{displaymath}
r'= r_{Si} = \rho_{max} \sin(\alpha_{Si} /2)
\end{displaymath}

The initial estimate of $(z_i - z_v)$ is $(z_{Si} - z_v)$, and
$\rho_{max}$ is determined using equations~\ref{equ:rhomax} and
\ref{equ:cos_zeta}. In this case

\begin{displaymath}
(z_i-z_{Si})/(z_i-z_v)=\alpha_{Si}/2\pi
\end{displaymath}

from which a better estimate of $z_i$ can be determined.

The whole procedure is repeated and rapidly converges so that the
value of $z$ changes by less than 0.1mm, typically after 5
iterations or less.

\section*{Acknowledgements}
This work was supported by the Science and Technology Facilities
Council (UK) Grant No. ST/V001027/1. The software package supporting
this article is published~\cite{ButGH} under the GNU General Public
License v3.0.  The author acknowledges useful discussions with
Yassid Ayyad, Daniel Bazin, Andreas Ceulemans, Matthew Fraser, Sean
Freeman, Liam Gaffney, Jack Henderson, Benjamin Kay, Marc Labiche,
Alison Laird, Robert Page, Oleksii Poleshchuk, Riccardo Raabe, David
Sharp, and Fredrik Wenander.

\end{document}